\documentclass{article}
\usepackage{spconf,amsmath,graphicx,bm,cite,setspace}
\usepackage{lipsum}
\usepackage{multirow}

\title{Semi-supervised and Transfer learning approaches for low resource sentiment classification}
\name{Rahul Gupta$^o$, Saurabh Sahu$^+$, Carol Espy-Wilson$^+$, Shrikanth Narayanan$^*$}
\address{$^o$Amazon.com \\
$^+$Speech Communication Laboratory, University of Maryland, College Park \\
$^*$Signal Analysis and Interpretation Lab, University of Southern California, Los Angeles}

\begin{document}
\ninept
\maketitle
\begin{spacing}{1.0}
\begin{abstract}
Sentiment classification involves quantifying the affective reaction of a human to a document, media item or an event.
Although researchers have investigated several methods to reliably infer sentiment from lexical, speech and body language cues, training a model with a small set of labeled datasets is still a challenge.
For instance, in expanding sentiment analysis to new languages and cultures, it may not always be possible to obtain comprehensive labeled datasets.
In this paper, we investigate the application of semi-supervised and transfer learning methods to improve performances on low resource sentiment classification tasks.
We experiment with extracting dense feature representations, pre-training and manifold regularization in enhancing the performance of sentiment classification systems.
Our goal is a coherent implementation of these methods and we evaluate the gains achieved by these methods in matched setting involving training and testing on a single corpus setting as well as two cross corpora settings.
In both the cases, our experiments demonstrate that the proposed methods can significantly enhance the model performance against a purely supervised approach, particularly in cases involving a handful of training data.

\end{abstract}


\section{Introduction}
\label{sec:intro}

Sentiment classification \cite{pang2008opinion} is classical problem associated with determining the attitude and affective reaction of a person to an event, document and/or media item.
A typical setting for training sentiment classification models involves  feature extraction on a labeled corpus, followed by a classifier training \cite{pang2002thumbs}.
Such an approach has seen fair amount of success in several sentiment classification tasks such as twitter sentiment classification \cite{jiang2011target}, movie reviews \cite{kennedy2006sentiment} and product reviews \cite{cui2006comparative}.
However, as the problem of sentiment classification expands to new arenas, such as new modes of expressions on social media, different languages and even different cultures, large amounts of labeled corpora may not be immediately available. 
A combination of transfer learning and semi-supervised methods can then be used to obtain better initial models, which can then be refined as more training data becomes available.
In this paper, we analyze performance trends with varying amount of labeled training data using three different semi-supervised and transfer learning methods: (i) learning feature representations on external data, (ii) model pre-training and, (iii) manifold regularization.
We perform experiments on a single corpus setting as well as in two cross-copora setting to evaluate the impact of these methods.
Through our experiments, we aim to investigate the applicability of these methods under different dataset conditions and present recommendations. 

{\bf Previous work}: Researchers have proposed numerous semi-supervised and transfer learning methods \cite{pan2010survey,zhu2006semi} with successful applications to reinforcement learning \cite{taylor2009transfer}, natural language understanding \cite{liang2005semi}, speech recognition \cite{yu2010active}, as well as sentiment analysis \cite{millersemi,goldberg2006seeing,calais2011bias}.
Within the purview of semi-supervised learning Miller et al. \cite{millersemi} use a label propagation technique, assigning sentiment labels to unlabeled sentences from a labeled neighbor.
Along similar lines, Goldberg et al. \cite{goldberg2006seeing} use graph regularization to use unlabeled data to improve classification using a linear support vector machine. 
However, we note that hard label propagation as performed by Miller et al. \cite{millersemi} may not always be appropriate and depends on feature representation for the sentences.
On the other hand, Goldberg et al. \cite{goldberg2006seeing} perform manifold regularization using a crafted similarity metric, which may not be generically applicable.
Other semi-supervised techniques make use of co-regularization \cite{sindhwani2008document}, active deep networks \cite{zhou2010active} and joint sentiment-topic detection \cite{lin2012weakly}.
Transfer learning approaches for sentiment classification include structural correspondence learning \cite{blitzer2007biographies} and spectral feature alignment \cite{pan2010cross}.
Transfer learning focuses on adapting features \cite{glorot2011domain} or aid model training \cite{tan2009adapting} on a related corpus to enhance performance on the problem of interest.
In our paper, we experiment with a combination of semi-supervised and transfer learning with a motivation towards coherent implementation of the techniques.

We assume a setting with a small set of labels available on the task of interest and experiment with learning a feature representation, initializing a classifier using pre-training and finally performing a semi-supervised optimization.
In order to learn feature representations we use sentence embedding using the doc2vec model \cite{le2014distributed}.
The doc2vec tends to cluster sentences with similar meaning together desirable for a discriminative classification setup.
We further hypothesize that the representations learnt using the doc2vec models are similar across datasets, therefore models pre-trained on an external dataset can be used for classification on the dataset at hand.
For the same reason, we also hypothesize that sentences from external dataset can be used for manifold regularization based semi-supervised learning.
We empirically test these hypotheses on a single corpus setting involving unlabeled data available on the dataset of interest, as well as two cross corpora settings which make use of data from an external source.
We demonstrate the success of semi-supervised and transfer learning methods, particularly in cases when a small amount of labeled data is available on the task of interest.
The gain in performance tends to decrease as more labeled data is made available.
We discuss the methodology in the next section, followed by the experimental setup in Section~\ref{sec:exp}.


\section{Methodology}
We make use of the following techniques for training a sentiment classification model: (i) learning feature representations on an external corpus, (ii) pre-training and (iii) manifold regularization.
Learning feature representations and pre-training on external corpus can be viewed as transfer learning methods, while the manifold regularization technique uses a mix of external and in-domain data to improve classification models, without the requirement of labels.
Therefore, it can be viewed as both, a transfer learning and a semi-supervised approach.
We discuss these techniques in more detail below.

\subsection{Learning feature representations}
\label{sec:feat_rep}
Given a dataset with a set of sentences, we extract a vector representation for every sentence using a doc2vec \cite{le2014distributed} model.
Doc2vec models provide a compact projection of the sentences by projecting the high dimensional feature space spanned by the n-grams in the language vocabulary \cite{le2014distributed}.
Learning feature representations on low resource datasets is challenging due to limited vocabulary coverage (for instance only a few n-grams can be observed on a small dataset).
Hence, it is possible to observe words during testing, which were not observed during learning the feature representations (these words are typically considered to be out of vocabulary words during testing).
Doc2vec model trained on a larger external corpus learns a representation for a large number of sentence formations, and the representations tend to cluster for sentences carrying similar semantic meanings. 
The clustering of semantically similar sentences is a desirable property for training a discriminative classifier.
We train the doc2vec model on Wikipedia articles consisting of approximately 4 million articles \cite{dai2015document}.
Note that we do not use any sentiment labels for the doc2vec model training and therefore the feature extraction is unsupervised.
We acknowledge that training Doc2vec models on in-domain data is desirable to avoid domain mismatch, however training these models typically requires a large amount of data, which is not possible in low resource classifications tasks. 

\subsection{Pre-training classification models}
After obtaining the feature representations using the doc2vec model, we initialize the classification model for low resource dataset by performing supervised training on a large external corpus.
We chose this external corpus to be closely associated with the task at hand, but the corpus may be collected with a different objective.
Note that the sentiment labels used to pre-train the classification models may carry different connotation for different datasets.
A second pass training is done on the in-domain data, and we aim to obtain a better model initialization using the external corpus with the assumptions that the definition of the sentiment labels is loosely associated for the external and in-domain datasets.

\subsection{Model training with manifold regularization}
Post feature extraction, we propose the application of manifold regularization to train a statistical model to use the labeled and unlabeled data resources. 
Manifold regularization was proposed by Belkin et al. \cite{belkin2006manifold} and adds a regularization penalty term to the supervised loss.
Given a set of labeled feature vectors $\bm x_i (i = 1,..,l)$, with a corresponding label $y_i$,  a choice of Reproducible Kernel Hilbert Space (RHKS) $\mathcal H_k$ and a loss function $V$, Belkin et al. \cite{belkin2006manifold} define the optimization problem in equation~\ref{eq:belkin} to yield a classifier function $f^*$ belonging to the space $\mathcal H_k$.
In the equation, $V(x_i, y_i, f)$ can be any loss function (e.g. mean squared error: $|y_i - f(x_i)|^2_2$, cross-entropy: $y_i \log f(x_i) + (1-y_i) \log (1-f(x_i))$).
$||f||_k^2$ is a regularization cost controlling the intrinsic structure of the classifier (e.g. L1 or L2 penalty) and $||f||_I^2$ is an additional smoothness loss controlling the complexity of the classifier along the distribution of the set of labeled and unlabeled data points (please refer to Section 2 in \cite{belkin2006manifold} for more details).
$\gamma_A$ and $\gamma_I$ are the hyper-parameters controlling the trade-off amongst various losses in the equation \ref{eq:belkin}.

\begin{equation}\label{eq:belkin}
f^* = \arg\min_{f \in \mathcal H_k} \frac{1}{l} \sum_{i=1}^l V(x_i, y_i, f) + \gamma_A ||f||_k^2 + \gamma_I ||f||_I^2
\end{equation}

For the purpose of our experiments, we learn a neural network as the function $f$, $V(x_i, y_i, f)$ is set to the cross-entropy loss and we use L2 regularization on the neural network weights as the loss $||f||_k^2$.
We set $||f||_I^2$ to the following value in equation~\ref{eq:manifold_loss}.

\begin{equation}\label{eq:manifold_loss}
||f||_I^2 = \sum_{i=1}^l \sum_{\substack{{x_i^u \in} \\ {\text{Neighborhood of } x_i}}} \frac{||f(\bm x_i) - f(\bm x_i^u)||^2_2} {||\bm x_i - \bm x_i^u||_2}
\end{equation}

The loss minimizes the Euclidean distance between the outputs for labeled instance $\bm x_i$:$f(\bm x_i)$  and a set of unlabeled data-points in the neighborhood of $\bm x_i$:$f(\bm x_i^u)$.
The loss is inversely weighted by the distance between $\bm x_i$ and $\bm x_i^u$, so that the loss $||f(\bm x_i) - f(\bm x_i^u)||^2_2$ carries a higher importance when $\bm x_i^u$ is closer to $\bm x_i$ in the local Euclidean vicinity. 
We hypothesize that this setup is particularly useful in the case of feature representations obtained from the doc2vec models.
Since a doc2vec model tends to cluster utterances with similar meaning together, penalizing the difference between model outputs for neighboring points is desired.
During optimization, we draw the unlabeled data from an external data in addition to in-domain unlabeled resources, if available. 
We optimize the loss using the SGD (Stochastic Gradient Descent) optimizer in Keras (a high-level neural networks API in Python \cite{chollet2015keras}).

We note that initiating with a coherent feature representation for the sentences is desired for the success of pre-training and manifold regularization techniques.
We derive a dense representation for various datasets using a single doc2vec model trained on a larger corpus.
This methodology is likely to yield consistent representations across datasets, where utterances carrying similar semantic connotations tend to have similar representations.

\section{Experiments}
\label{sec:exp}
We perform evaluation of semi-supervised methods under two settings: (i) single corpus setting and, (ii) cross corpora setting.
In the single corpus setting, semi-supervised methods are applied to in-domain data while in the cross corpora setting, we use related data available from other corpora to improve performance on a task at hand.
Next, we describe these experiments in detail.

\subsection{Single corpus setting}
\label{sec:single_corp}
This experiment addresses the cases where a lot of data is available for the task of interest, however only a partial set of data is annotated.
We use the Sentiment140 corpus \cite{go2009twitter} for this experiment. 
The corpus consists of $\sim$1.6M tweets marked with a positive or a negative sentiment.
We randomly and equally split the data into a training, development and testing partition.
We extract the feature representations for the tweets using the doc2vec model described in Section~\ref{sec:feat_rep}.
This is followed by semi-supervised training using the available in-domain unlabeled dataset.
We do not perform pre-training in this experiment, as we assume the availability of partially annotated samples only from a single corpus.
We evaluate the performance of the resulting models against two fully supervised training baselines: (i) a model trained using only the available set of labeled data and, (ii) a model trained with the assumption that labels are available on the entire training set. 
The baselines are also trained on the doc2vec representations and serve as lower and upper bounds on the performance of the semi-supervised loss.
The hyper-parameters of the neural network $f$ (number of nodes in the hidden layer), $\gamma_A$ and $\gamma_I$ are tuned on the development set. 
Note that supervised baselines essentially set $\gamma_I$ to 0.
We perform multiple evaluations with an increasing proportion of labeled data-points provided during model training.
We present our results in the next section.

\subsubsection{Results}

\begin{figure}
\includegraphics[scale=0.45]{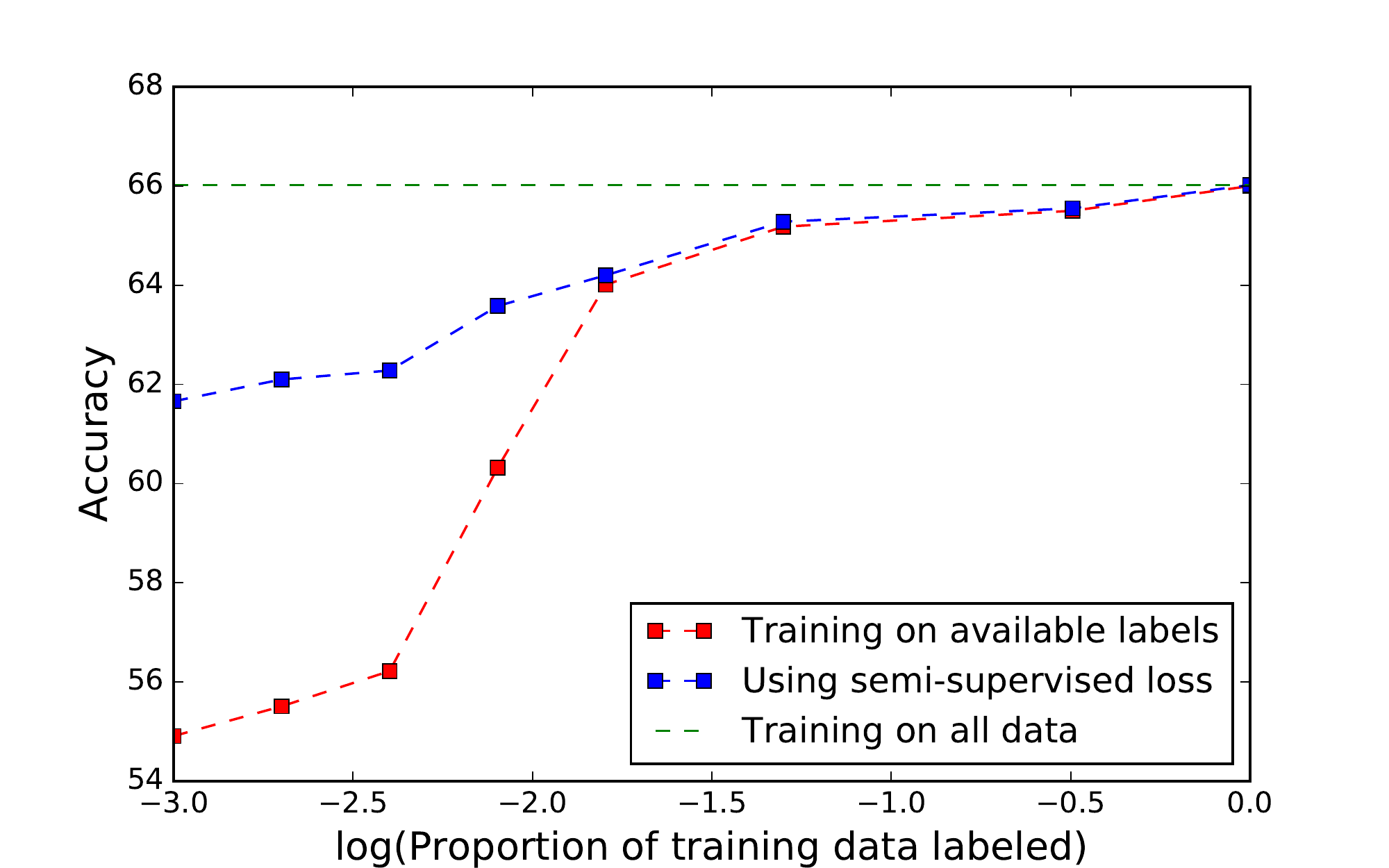}
\caption{Results on the Sentiment140 corpus. 
$\log$ is to the base 10 (we use $10^x$ proportion of all datapoints, where $x$ are values denoted by the x-axis. For instance when $x=-2$, we use 1\% of labeled data for training).
The line used to depict training on all data is only for comparison and should not be linked to x-axis.}
\label{fig:single_results}
\end{figure}

Figure~\ref{fig:single_results} presents the results using the two baselines and training using manifold regularization.
From the results, we observe that semi-supervised training using manifold regularization outperforms the purely supervised approach, particularly when a smaller fraction of training data is assumed to be labeled. 
This is expected as the unlabeled data helps regularize the model outputs along the manifold on which the data lie in the feature space.
We also performed another experiment by training the doc2vec models on the entire training and development set partition ($\sim$1M tweets).
However these models consistently under-performed the doc2vec representations obtained from the Wikipedia corpus (the best performance achieved by in-domain doc2vec representation was 59.2\% when labels on the entire training data were made available).
The Wikipedia corpus consists of ~4M articles, which yield better doc2vec representations versus training on smaller set of ~1M tweets with a few words in each tweet.
In the next section, we discuss a more challenging case of applying training using resources from external corpora.

\subsection{Semi-supervised learning: Cross corpora setting}
In the previous experiment, we investigate methods to improve classification performance under a matched setting, where partially labeled training data and testing data are drawn from the same corpus.
A separate setting could involve the availability of limited training data in a corpus of interest, however a larger set of a labeled training data may be available on a related external corpus. 
We apply the proposed methods on two corpora described below.

{\bf UCI sentiment labeled sentences data set}: The UCI sentiment labeled sentences data set \cite{kotzias2015group} consists of 3000 sentences accumulated from amazon.com, yelp.com and imbd.com.
They are labeled as either `positive' or `negative'.
We randomly split the data into half, using one partition for training and the other one for testing.

{\bf Movie review dataset}:
The movie review dataset \cite{pang2004sentimental} consists of 2000 samples consisting of movie reviews with multiple sentences. 
Note that the average length of these reviews is longer than the Sentiment140 corpus and the UCI sentiment labeled sentences data set.
The data is partitioned into a training and a testing set consisting of 1000 samples each. 
Each movie review is again labeled as `positive' or `negative'.

Given the two datasets, we initially extract feature representations from the doc2vec model trained on Wikipedia corpus.
During model pre-training, we use all of the Sentiment140 corpus to obtain weights for the neural network.
Finally, during model training, we assume that the in-domain training set is partially labeled and is used to compute the cross entropy loss ($V(x,y,f)$).
For computing the manifold regularization loss ($||f||_I^2$), unlabeled data is sourced from in-domain unlabeled data and the Sentiment140 corpus.
Since the datasets of interest contain a few thousand samples, we hypothesize that regularization using external data can help achieve better model generalization.
We use the baseline training methodologies specified in Section~\ref{sec:single_corp} and perform multiple evaluations for the proposed methods in performed by increasing the quantity of available labels on the training set. 
In order to independently estimate the effects of pre-training and model training we conduct the following set of experiments apart from the two baselines: (i) model pre-training + supervised training on available data, (ii) model training based on manifold regularization with no pre-training and, (iii) pre-training followed by training using manifold regularization.
Note that we do not use a development set to tune the hyper-parameters of our model as we assume a limited availability of labeled samples during training.
We chose the model configuration as the one that performed best on the Sentiment 140 corpus.


\begin{figure}[t]
\includegraphics[scale=0.35]{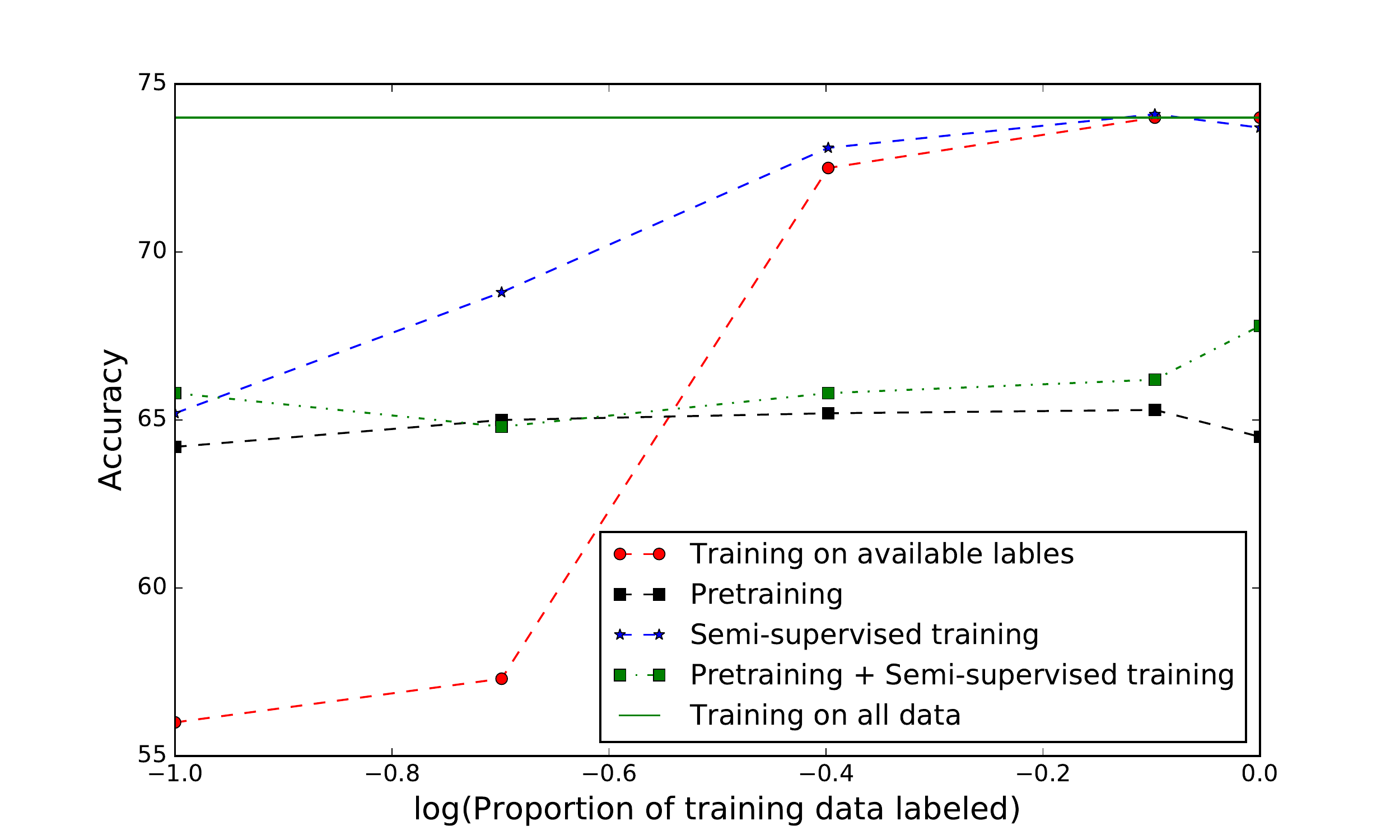}\\
\includegraphics[scale=0.35]{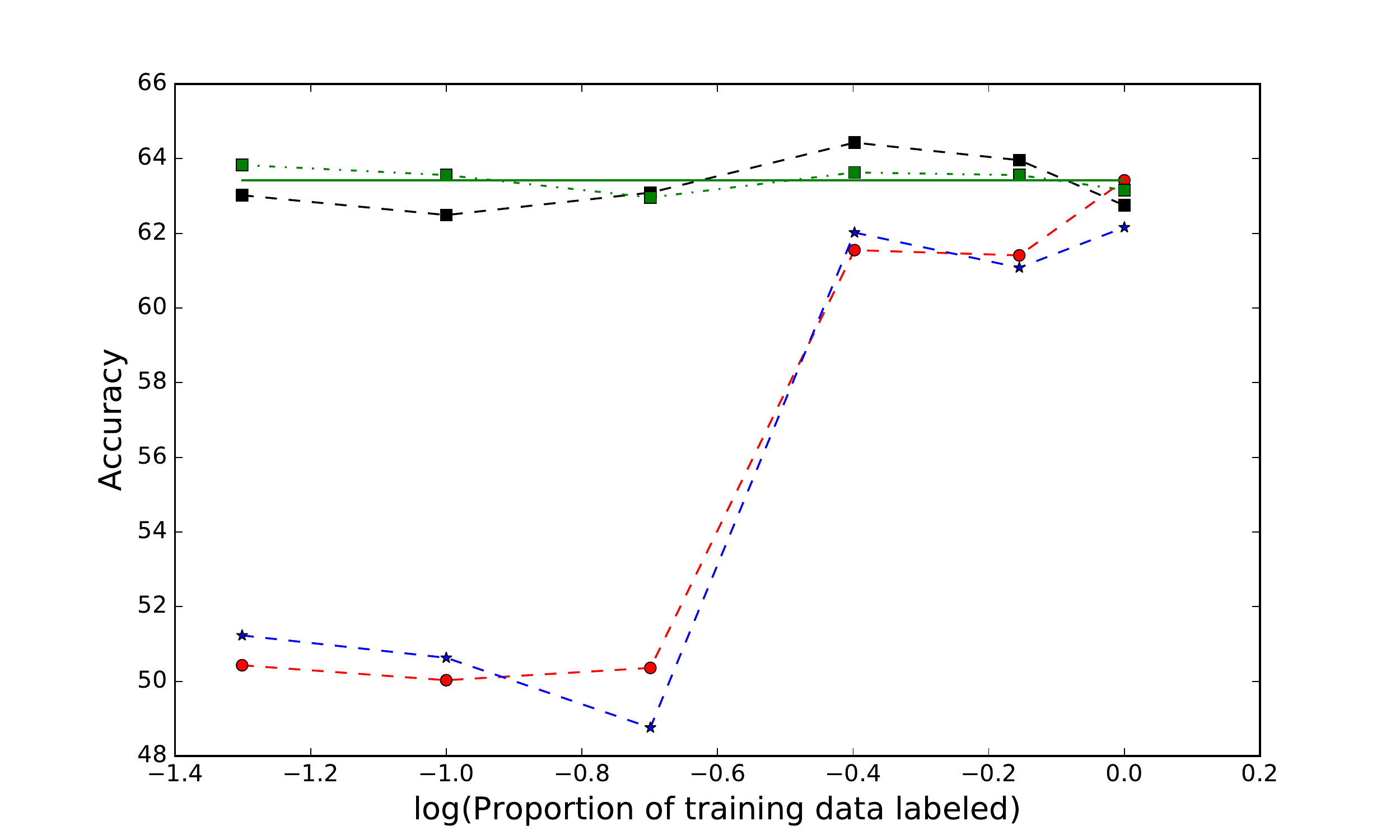}
\caption{Cross-corpora study: Results on the movie review corpus (top) and the UCI sentiment corpus (bottom).}
\label{fig:cross_results}
\end{figure}

\subsubsection{Results}
Figure~\ref{fig:cross_results} presents the results on the two datasets.
From the results, we observe that the model pre-training works particularly well with small amounts of training data. In the case of the UCI sentiment data, the pre-trained model works as good as supervised model training on all the 1500 data samples. However, in the case of the movie review dataset, pre-training provides an advantage only with smaller amounts of training data and results do not improve with availability of more in-domain training data. 
We also observe that the semi-supervised regularization outperforms baseline supervised training in the case of movie review dataset.
On the other hand, manifold regularization does not provide gains in the case of UCI sentiment dataset.
In case of availability of small amounts of labeled data, the purely supervised approach performs close to chance accuracy.
We do not expect manifold regularization to provide improvement in this case as regularization using unlabeled data is performed using labeled data predictions, which are unreliable in this case.
We also observe a quick saturation of performance as we add more data, another case when manifold regularization does not provide any further gains.

\begin{figure}[t]
\includegraphics[scale=0.35]{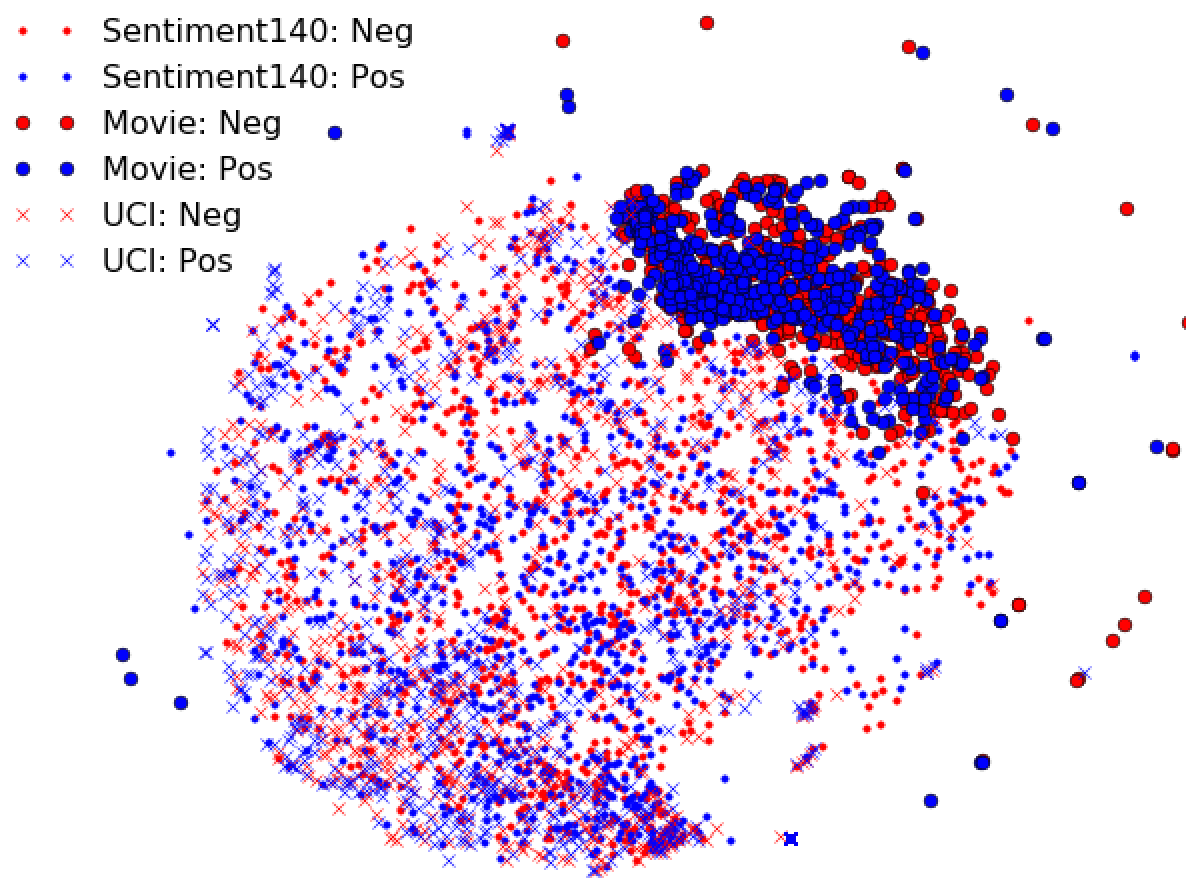}
\caption{t-SNE plots for UCI sentiment dataset, movie review dataset and a randomly sampled set from the Sentiment140 corpus.}
\label{fig:tsne}
\end{figure}

To further understand the impact of pre-training and manifold regularization in a cross corpora setting, we plot the t-Stochastic Neighbor Embedding (t-SNE) \cite{maaten2008visualizing} plots for the three datasets in Figure~\ref{fig:tsne}.
We plot 2000 randomly selected Sentiment-140 representation and all of movie review and UCI sentiment datasets on a t-SNE model learnt on all of the data combined.
From the figure, we observe that the distribution of the movie review dataset is different from the other two datasets, explaining that pre-training on Sentiment140 corpus does not achieve the same level of performance as a supervised model trained on all of the training data.
On the other hand, the UCI sentiment dataset follows a similar distribution based on the t-SNE projections and hence pre-training is expected to yield matched models.
In case of manifold regularization, we use a mix of in-domain labeled data and Sentiment140 corpus.
Although the t-SNE plots suggest that the UCI sentiment and the Sentiment140 corpus datapoints follow similar distributions, no gains are observed due to quick saturation of performance from a chance model.
Since pre-training followed by in-domain training does not lead to performance improvement as seen for movie review dataset (Figure~\ref{fig:cross_results} (top)), we recommend its implementation based on a data distribution analysis as done using the t-SNE plot.
This analysis is important as post pre-training on a large dataset, in-domain training does not improve performance beyond the one achieved by the pre-trained model.

\section{Conclusion}
Sentiment expression is universal across languages and cultures.
Scaling it to new arenas may need to overcome the data sparsity challenges.
We explore semi-supervised and transfer learning approaches to improve performance on low resource sentiment classification tasks.
Initially, we learn dense representations for sentences using a doc2vec model, followed by experimentation with pre-training and manifold regularization.
We observe gains using the proposed methods on a single corpus setting as well as two cross corpora settings.
In particular when a handful of training data is available, the improvements are significant over a purely supervised approach.

In the future, we aim to extend the same study to transfer settings across tasks with different but related output labels (e.g. learning a sentiment classification system using an external emotion corpora).
We also aim to test other forms of semi-supervised learning methods involving domain adaptation and feature transformation \cite{glorot2011domain}.
Furthermore, researchers have proposed alternate methods for sentence representations with different motivations \cite{karpathy2014deep}.
One can carry out investigations regarding the impact of these representation on the model performances.
Finally, this study can be extended to a multi-task setting where transfer of learning can be performed across tasks apart from across datasets.

\end{spacing}

\newpage

\begin{spacing}{1.0}
\bibliographystyle{IEEEtran}
\bibliography{strings}
\end{spacing}
\end{document}